\documentclass[aps,prx,onecolumn,eqsecnum]{revtex4-2}  
\usepackage{amsmath,amssymb,bm,bbm} 
\usepackage{graphicx}  
\usepackage{color} 
\usepackage{tikz}
\usepackage{booktabs}
\usepackage{array}

\usepackage[papersize={8.5in,11in}]{geometry}
\usepackage[colorlinks=true]{hyperref}  
\usepackage[mathscr]{euscript}
\usepackage[section]{placeins}
\DeclareMathAlphabet{\mathpzc}{OT1}{pzc}{m}{it}
\hypersetup{
    bookmarks=true,         
    unicode=false,          
    pdftoolbar=true,        
    pdfmenubar=true,        
    pdffitwindow=false,     
    pdfstartview={FitH},    
    pdfsubject={},   
    pdfcreator={},   
    pdfproducer={}, 
    pdfkeywords={} {} {}, 
    pdfnewwindow=true,      
    colorlinks=true,       
    linkcolor=magenta, 
    citecolor=blue,        
    filecolor=magenta,      
    urlcolor=blue           
} 

\usepackage{amsfonts}
\usepackage{upgreek}
\usepackage{slashed}
\usepackage{latexsym}

\usepackage{tcolorbox}

\newcommand{\beq}{\begin{equation}}
\newcommand{\eeq}{\end{equation}}
\def\bea{\begin{eqnarray}}
\def\eea{\end{eqnarray}}

\newcommand{\vi}{{\boldsymbol{i}}}

\newcommand{\nus}{\nu_{\rm spin}}

\begin{document}


\title{Constraints on the volume enclosed by Fermi surfaces}

\author{Subir Sachdev}
\affiliation{Department of Physics and Leinweber Institute for Theoretical Physics, Harvard University, Cambridge MA 02138, USA}

\date{\today
\\
\vspace{0.2in}
\begin{center}
{\tt For the Special Centenary Issue of the Indian Journal of Physics, {\it Recent Trends in Condensed Matter Physics}}
\end{center}
\vspace{0.2in}
}

\begin{abstract}
A fundamental property of a Fermi liquid (FL) is that the volume enclosed by the Fermi surface is determined by $\rho$, the total density of electrons.
In a fractionalized Fermi liquid (FL*), the volume enclosed by the Fermi surface is determined by the same expression after $\rho$ has been shifted by a {\it quantized\/} amount, most commonly to $\rho -1$ (we assume spin rotation invariance, for simplicity). The FL* shift is related to the quantized symmetry fractionalization data of a quantum spin liquid, whose fractionalized spinon excitations are present alongside the Fermi surface quasiparticles. In this review article, we establish the FL* Fermi volume by generalizing the original Luttinger proof for FL to parton gauge theories of the quantum spin liquid and FL*. We also discuss Oshikawa's flux insertion arguments for FL, and their extension to FL*, including a discussion of gapless spin liquids.
\end{abstract}

\maketitle

\tableofcontents

\section{Introduction}
\label{sec:intro}

In 1960, Luttinger \cite{Luttinger60} proved that Fermi liquids (FL) obey 
\begin{align}
\frac{\mbox{volume enclosed by the Fermi surface}}{\mbox{volume of the Brillouin zone}} = \frac{\mbox{$\rho$ mod 2}}{2} \quad \mbox{in FL,}\label{lutt1}
\end{align}
where $\rho$ is the total electron density, and we are assuming spin-rotation symmetry. (The volume enclosed by a Fermi surface is counted with a negative sign if it encloses quasihole states; this is the convention used for the hole pockets discussed below.) Luttinger's argument was based upon an analysis of the analytic structure of the electron's Green's function, and implicitly assumed the validity of an all orders expansion from the free fermion state; we will review this argument in Section~\ref{sec:gLutt}. Oshikawa \cite{MO00} placed Luttinger's argument on a non-perturbative basis by presenting a flux-piercing argument which has a modern interpretation as an `anomaly' associated with translations and charge conservation: Oshikawa's argument is reviewed in Section~\ref{sec:fl}. 

The idea of a `fractionalized Fermi liquid' (FL*) was introduced in Refs.~\cite{TSSSMV03,TSSSMV04}, as a metallic state in which the Fermi surface did {\it not\/} obey the Luttinger constraint. In the simplest case, the fractional volume enclosed by the Fermi surface in FL* is $(\rho-1)/2$. As stated in Ref.~\cite{TSSSMV03}:
\begin{quote}
``The primary purpose of this paper is to show that there
exist nonmagnetic, metallic states (FL*) in dimensions
$d \geq 2$ with a Fermi surface of ordinary $S=1/2$, charge
$e$, sharp quasiparticles, enclosing a volume
\begin{equation}
\mathcal{V}_{\mathrm{FL}^*} = \mathcal{K}_d \left[ (\rho_a - 1) (\mathrm{mod 2})\right], \nonumber
\end{equation}
over a finite range of parameters.''
\end{quote}
(Here, $\rho_a$ is the total electron density, $\mathcal{K}_d = (2 \pi)^d /(2 v_0)$ is a phase space factor, $v_0$ is the volume of the unit cell, and the (mod 2) accounts for filled bands which we ignore here.) So Eq.~(\ref{lutt1}) is replaced by
\begin{align}
\frac{\mbox{volume enclosed by the Fermi surface}}{\mbox{volume of the Brillouin zone}} = \frac{\mbox{$(\rho-1)$ mod 2}}{2} \quad \mbox{in FL*.}\label{lutt2}
\end{align}
The central point \cite{TSSSMV03,TSSSMV04,APAV04,Powell05,Coleman05,Qi10,SSMetlitskiPunk12,Bonderson16,SenthilElse21,QPMbook,Seiberg23,Meng26} was that it was possible to satisfy Oshikawa's anomaly-argument by combining the anomaly of a Fermi surface, which contributes an amount equivalent to a density $\rho-1$, with that of a fractionalized spin liquid, which contributes a {\it quantized\/} amount equivalent to a density 1.
It is trivially possible to shift $\rho$ by an even integer by adding and removing filled bands, and the novelty is the shift in FL* by an odd integer. Note that the shift in density cannot be arbitrary and is quantized by the rigid structure of the spin liquid under symmetry transformations. These anomaly arguments will be discussed further in Section~\ref{sec:oshikawa}, including the important case of gapless spin liquids in Section~\ref{sec:dqcp}.

It is worth emphasizing an important aspect of this quantized shift. In an Anderson lattice model, with a finite Hubbard $U$ for the $d$ electrons and a finite hybridization between the $d$ and conduction electrons, we can postulate a FL* state with a spin-liquid sector of $d$ electrons, and a Fermi surface of gauge-neutral electron-like quasiparticles from the conduction electrons enclosing the volume in Eq.~(\ref{lutt2}). At finite hybridization, these quasiparticles need not have purely conduction-orbital character, and the average $d$-orbital occupancy need not be 1. Nevertheless, provided the system remains in the same FL* phase, the shift in Eq.~(\ref{lutt2}) remains quantized at 1. The invariant is the {\it symmetry-fractionalization data of the spin liquid}, represented in the parton construction by an odd background gauge charge per unit cell, rather than the average $d$-orbital occupancy. 
Section~\ref{sec:gLutt} applies the Luttinger-style arguments to parton gauge theories, and demonstrates how Eq.~(\ref{lutt2}) follows from the quantization of the background gauge charge.

We note a recent review \cite{ROPP25} discussing FL* as a theory of the pseudogap metal phase of the hole-doped cuprates. The strongest evidence for the FL* description of the cuprates is provided by:
\begin{itemize}
\item Observation of coherent interlayer transport in angle-dependent magnetoresistance \cite{Ramshaw22,Yamaji24}, including the Yamaji effect \cite{Yamaji24}, is consistent with the FL* theory \cite{Zhao_Yamaji_25,FuChun25}. For the single-band model with electron density $\rho = 1-p$, there are 4 hole pockets, and each should enclose a fractional area $p/8$ by Eq.~(\ref{lutt2}): this value is close to the $1.3\%$ fractional area deduced from the Yamaji analysis for $p=0.1$ \cite{Yamaji24}, for which $p/8 = 1.25\%$.
\item Local density of states measurements by scanning tunneling microscopy in the underdoped superconducting state \cite{Seamus26} are very well fit by a theory of the superconductor obtained by lowering temperature on FL* \cite{Christos:2023oru}. These measurements show fingerprints of the spinons of the quantum spin liquid of FL*, which hybridize with the Bogoliubov quasiparticles in the superconducting state.
\end{itemize}
The second observation above, along with some earlier indications \cite{ROPP25}, requires FL* with a gapless spin liquid, similar to the deconfined quantum critical states discussed in Section~\ref{sec:dqcp}.

\FloatBarrier

\section{Luttinger constraints for fractionalized states}
\label{sec:gLutt}

The conventional proof of the Luttinger theorem for a Fermi liquid \cite{Luttinger60} rests on an all orders expansion from a free fermion state, and shows that the volume enclosed by the Fermi surface depends only on the density of fermions, and is independent of the strength of interactions. The proof involves certain assumptions on the structure of the perturbative expansion, which will be specified below.

In this section, we will present Luttinger's analysis in a more general context \cite{Powell05,Coleman05}. Rather than starting from a free fermion state, we will start with a simple saddle-point of a parton theory of a fractionalized metallic state. The perturbative expansion from this saddle point will necessarily involve emergent gauge fields, but this does not preclude application of Luttinger's methods as long as the perturbative expansion does not lead to an instability of the saddle point (which we assume). This generalized analysis also leads to constraints on the volume enclosed by the Fermi surface, but the constraints are distinct from those of Luttinger. 

As an example of a parton theory, consider the parton decomposition of an electron $c_\alpha$ into a spinful, neutral boson $b_\alpha$, and 
a spinless fermionic holon, $h$, of charge $e$. 
\begin{align}
    c_{\vi \alpha} = b_{\vi \alpha} h_\vi^\dagger\,. \label{eq:chb}
\end{align}
Associated with this parton decomposition is an emergent U(1) gauge invariance, and a constraint 
\beq
b_{\vi \alpha}^\dagger b_{\vi \alpha}^{\vphantom\dagger} + h_{\vi}^\dagger h_\vi^{\vphantom\dagger} = 1\,, \label{ll1}
\eeq
which fixes a unit background gauge charge on each site. Note that the value of the r.h.s. of Eq.~(\ref{ll1}) is a {\it rigid\/} property of the spin liquid which does not change upon doping. It is this background gauge charge which is responsible for the quantized Berry phase of $\pi$ for a vison encircling each site in a $\mathbb{Z}_2$ spin liquid \cite{RJSS91,SSMV99, HPS11}, and the quantized monopole Berry phase \cite{NRSS90,BS07,PCAS16} for the $\mathbb{CP}^1$ spin liquid (Section~\ref{sec:dqcp}). And as we will see in our discussions, 
\emph{it is also this quantized background gauge charge which is responsible for the quantized shift in Eq.~(\ref{lutt2}).}

Distinct from the U(1) gauge invariance, we also have a global U(1) symmetry associated with the conservation of electron density $\rho = 1-p$. In the parton formulation, an electron density 1 has been absorbed by the background gauge charge of the spin liquid, and the global density constraint is on the density of holons
\beq
 - \left\langle h_{\vi}^\dagger h_\vi^{\vphantom\dagger} \right\rangle = \rho - 1 \label{ll2}\,.
\eeq
The $-1$ on the r.h.s. of Eq.~(\ref{ll2}) is therefore directly connected to the 1 on the r.h.s. of Eq.~(\ref{ll1}). (The minus sign on the l.h.s. arises because $\rho$ measures the density of electrons, while $h$ counts holes.)

To extend this first example to an Anderson lattice model with a finite Hubbard $U$ on the $d$ orbitals, we retain the doubly occupied local state by introducing a spinless fermionic doublon $D_\vi$. For a $d$ orbital, the four physical local states are represented by
\beq
|0\rangle_\vi = h_\vi^\dagger |\mathrm{vac}\rangle\,, \qquad
|\alpha\rangle_\vi = b_{\vi\alpha}^\dagger |\mathrm{vac}\rangle\,, \qquad
|\uparrow\downarrow\rangle_\vi = D_\vi^\dagger |\mathrm{vac}\rangle\,,
\eeq
where $|\mathrm{vac}\rangle$ is the auxiliary parton vacuum. The projected decomposition (\ref{eq:chb}) must then be extended to include transitions involving the doublon. All three parton species carry unit emergent U(1) gauge charge, and the local constraint becomes
\beq
\sum_\alpha b_{\vi\alpha}^\dagger b_{\vi\alpha}
+ h_\vi^\dagger h_\vi + D_\vi^\dagger D_\vi = 1\,. \label{ll1finiteU}
\eeq
The physical $d$-electron number is
\beq
n_{d,\vi} = \sum_\alpha b_{\vi\alpha}^\dagger b_{\vi\alpha}
+ 2 D_\vi^\dagger D_\vi
= 1 - h_\vi^\dagger h_\vi + D_\vi^\dagger D_\vi\,. \label{ndfiniteU}
\eeq
Thus charge fluctuations change the average $d$-orbital occupancy through the holon and doublon densities while leaving the unit background gauge charge unchanged. Under the global electron-number U(1), the holon and doublon have charges $-1$ and $+1$, respectively, while the spinons are neutral. For an Anderson lattice with one $d$ orbital per unit cell and conduction-electron density $\rho_c$, the total electron density therefore obeys
\beq
\rho - 1 = \rho_c - \langle h_\vi^\dagger h_\vi\rangle
+ \langle D_\vi^\dagger D_\vi\rangle\,. \label{ll2Anderson}
\eeq
This is the conserved charge to which the generalized Luttinger analysis below applies. Under the assumptions of that analysis, if the charged Fermi-surface contribution is exhausted by gauge-neutral electron-like quasiparticles, their total signed Fermi volume per spin satisfies Eq.~(\ref{lutt2}). The subtraction is independent of the noninteger value of $\langle n_d\rangle$. The local constraint alone does not establish the existence or stability of this FL* phase; we assume that the charge fluctuations preserve its spin-liquid sector and symmetry-fractionalization data.

As a third example of a parton theory, consider the square lattice SU(2) spin liquid \cite{Christos:2023oru} which has neutral fermionic spinons $f_\alpha$, and a doublet $(B_1, B_2)$ of spinless, charge $e$ bosons. In the parton formulation of Ref.~\cite{LeeWen96}, the analog of the gauge charge constraint in Eq.~(\ref{ll1}) is
\beq
f_{\vi \alpha}^\dagger f_{\vi \alpha}^{\vphantom\dagger} + B_{1\vi}^\dagger B_{1\vi}^{\vphantom\dagger} - B_{2\vi}^\dagger B_{2\vi}^{\vphantom\dagger}  = 1\,, \label{ll3}
\eeq
and there is a similar constraint in the rotor formulation of Ref.~\cite{HermeleHoneycomb}. The r.h.s. of Eq.~(\ref{ll3}) is quantized by the SU(2) gauge symmetry. Now the global electron number symmetry leads to a constraint on the density of bosons
\beq
 - \left\langle B_{1\vi}^\dagger B_{1\vi}^{\vphantom\dagger} \right\rangle - \left\langle B_{2\vi}^\dagger B_{2\vi}^{\vphantom\dagger} \right\rangle = \rho - 1 \label{ll4}\,,
\eeq
in contrast to the density of fermions in Eq.~(\ref{ll2}).

\begin{figure}
\centering
\includegraphics[width=5.5in]{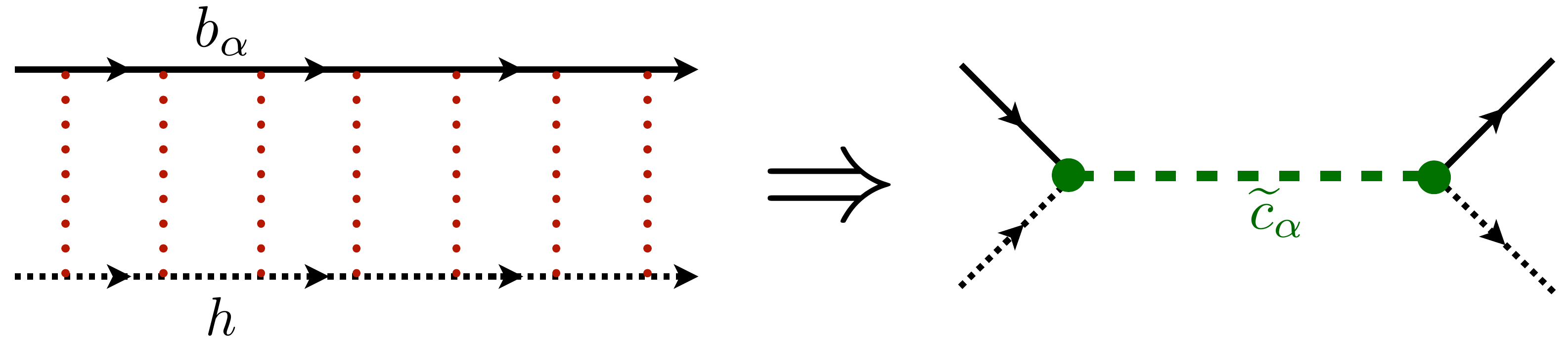}
\caption{$T$-matrix of holon-spinon scattering in the partons of Eq.~(\ref{eq:chb}) realized by an intermediate $\widetilde{c}_\alpha$ bound state. The green dot is the Yukawa coupling.}
\label{fig:feshbach}
\end{figure}

Next, we follow the analysis of Ref.~\cite{Powell05} to determine the low energy consequences of the microscopic constraints associated with electron number conservation given in Eqs.~(\ref{ll2}) and (\ref{ll4}), or Eq.~(\ref{ll2Anderson}) for the Anderson lattice. Any renormalized theory must have a global U(1) electron number symmetry with a net charge matching the appropriate microscopic constraint.
We assume that the renormalized theory is well described by a set of fields $F_s$ carrying charges $q_s$ under this global U(1) symmetry, 
so that the theory is invariant under
\beq
F_s \rightarrow F_s e^{i q_s \theta} \label{ll5}\,.
\eeq
The $F_s$ fields can be fermions or bosons, and will generally always include the parton fields of the microscopic theory. However, there can also be additional fields,
associated with `molecular' bound states of the partons. These bound states are associated with a pole in a parton-parton scattering $T$-matrix: such a pole can be replaced by the propagator of 
a canonical fermion or boson field representing the `molecule', along with a coupling allowing decay of the `molecule' into the `atomic' partons: see Fig.~\ref{fig:feshbach}.
With such a representation, it is possible to systematically treat the situation with finite densities of partons and their bound states \cite{Powell05,NishidaSon1,NishidaSon2,NikolicSS}. This process is similar to the `two channel' approach to the Feshbach resonance \cite{Feshbach58,Feshbach62,Feshbach08}, and effective field theories in nuclear physics \cite{Kaplan97}.
In particular, we include a `molecular' fermionic field which has the same quantum numbers as the electron, $\widetilde{c}_\alpha$, as a renormalized quasiparticle obtained as a bound state of a spinon and a holon \cite{RKK07,RKK08,Qi10}. The complete set of low energy fields is difficult to specify in advance for a particular Hamiltonian, but {\it a posteriori} the requirement is that the Luttinger-type analysis is well behaved, as we will clarify below. An explicit derivation of such a renormalized theory for the square lattice is provided by the ancilla method \cite{ROPP25,Mascot22}.

In a perturbative Feynman graph analysis \cite{Potthoff04}, the renormalized theory is characterized by corresponding renormalized Green's functions $G_s ({\bm p}, i \omega) $ and 
self energies $\Sigma_s ({\bm p}, i \omega) $ (assumed diagonal in the field index for convenience) which are related by a Luttinger-Ward functional $\Phi_{LW} [G_s]$ 
\beq
\Sigma_s ({\bm p}, i \omega)= \varsigma_s \frac{ \delta \Phi_{LW}}{\delta G_s ({\bm p}, i \omega)}\,, \label{flt23}
\eeq
where $\varsigma_s = -1 (+1)$ for fermions (bosons).
An important property of the Luttinger-Ward functional, following from the global U(1) symmetry in Eq.~(\ref{ll5}), is its invariance under frequency shifts
\beq
\Phi_{LW} \left[ G_s({\bm p}, i \omega + i q_s \omega_0) \right] = \Phi_{LW} \left[ G_s ({\bm p}, i \omega) \right]\,, \label{flt24}
\eeq
for any fixed $\omega_0$. Here, we are regarding $\Phi_{LW}$ as a functional of two sets of distinct functions $f_{s1,s2} (\omega)$, with $f_{s1} (\omega) \equiv G_s ({\bm p}, i \omega + i q_s \omega_0)$ and  $f_{s2} (\omega) = G_s ({\bm p}, i \omega) $, and $\Phi_{LW}$ evaluates to the same value for these two functions.

In terms of the renormalized theory, the microscopic constraints in Eqs.~(\ref{ll2}), (\ref{ll4}), and (\ref{ll2Anderson}) can be evaluated by computing the conserved global U(1) density:
\beq
\rho - 1 = -\frac{1}{\mathcal{V}} \sum_{\bm p}   \sum_s \varsigma_s q_s d_s \int_{-\infty}^{\infty}\frac{d \omega}{2 \pi} G_s ({\bm p}, i \omega) e^{i \omega 0^+} \,, \label{flt20}
\eeq
where $\mathcal{V}$ is the number of lattice sites, and $d_s$ is the spin degeneracy of particle $s$.
Now we can proceed along exactly the lines described by Luttinger \cite{Luttinger60,QPMbook}. We insert into Eq.~(\ref{flt20}) the following identity following directly from Dyson's equation
\beq
 G_s ({\bm p}, i \omega) =  i \frac{\partial}{\partial \omega} \ln \left[ G_s ({\bm p}, i \omega) \right]  - i G_s ({\bm p}, i \omega) \frac{\partial}{\partial \omega} \Sigma_s ({\bm p}, i \omega) \,. \label{flt21}
\eeq
Then we can write
\beq
\rho-1 = \mathcal{I}_{1p} + \mathcal{I}_{1z} + \mathcal{I}_2 \label{ll10}
\eeq
where
\bea
\mathcal{I}_{1p} &=&  - \left. \frac{i}{\mathcal{V}} \sum_{\bm p} \sum_s \varsigma_s q_s d_s \int_{-\infty}^{\infty}\frac{d \omega}{2 \pi}  \frac{\partial}{\partial \omega} \ln \left[ G_s ({\bm p}, i \omega) \right] e^{i \omega 0^+} \right|_{ \mbox{only from poles of $G_s$}} \nonumber \\
\mathcal{I}_{1z} &=&  - \left. \frac{i}{\mathcal{V}} \sum_{\bm p} \sum_s \varsigma_s q_s d_s \int_{-\infty}^{\infty}\frac{d \omega}{2 \pi}  \frac{\partial}{\partial \omega} \ln \left[ G_s ({\bm p}, i \omega) \right] e^{i \omega 0^+} \right|_{ \mbox{only from zeros of $G_s$}} \nonumber \\
\mathcal{I}_2 &=&  \frac{i}{\mathcal{V}} \sum_{\bm p} \sum_s \varsigma_s q_s d_s \int_{-\infty}^{\infty}\frac{d \omega}{2 \pi} G_s ({\bm p}, i \omega) \frac{\partial}{\partial \omega} \Sigma_s ({\bm p}, i \omega)\,.
\eea
The $\mathcal{I}_{1p}$ term contains the conventional contribution of the Fermi surface where $G_s^{-1} ({\bm p}, \omega=0) = 0$, while the $\mathcal{I}_{1z}$ is often called the contribution of the ``Luttinger surface'' where $G_s ({\bm p}, \omega=0) = 0$.

Important ingredients in the conventional Luttinger arguments are that $\mathcal{I}_{1z} = 0$ and $\mathcal{I}_2 = 0$. It is assumed that the perturbation theory is stable, there are no poles in the self energy $\Sigma_s$, and consequently no zeros in $G_s$, and so $\mathcal{I}_{1z}$ vanishes. The vanishing of $\mathcal{I}_2$ follows after taking the $\omega_0$ derivative of Eq.~(\ref{flt24}), and integrating by parts in frequency. Then the conventional Luttinger area of $\rho/2$ follows by evaluating $\mathcal{I}_{1p}$ from the electron quasiparticle pole on the Fermi surface, with the l.h.s. in Eq.~(\ref{ll10}) equal to $\rho$ in this case.

\begin{figure}
\centering
\includegraphics[width=5.5in]{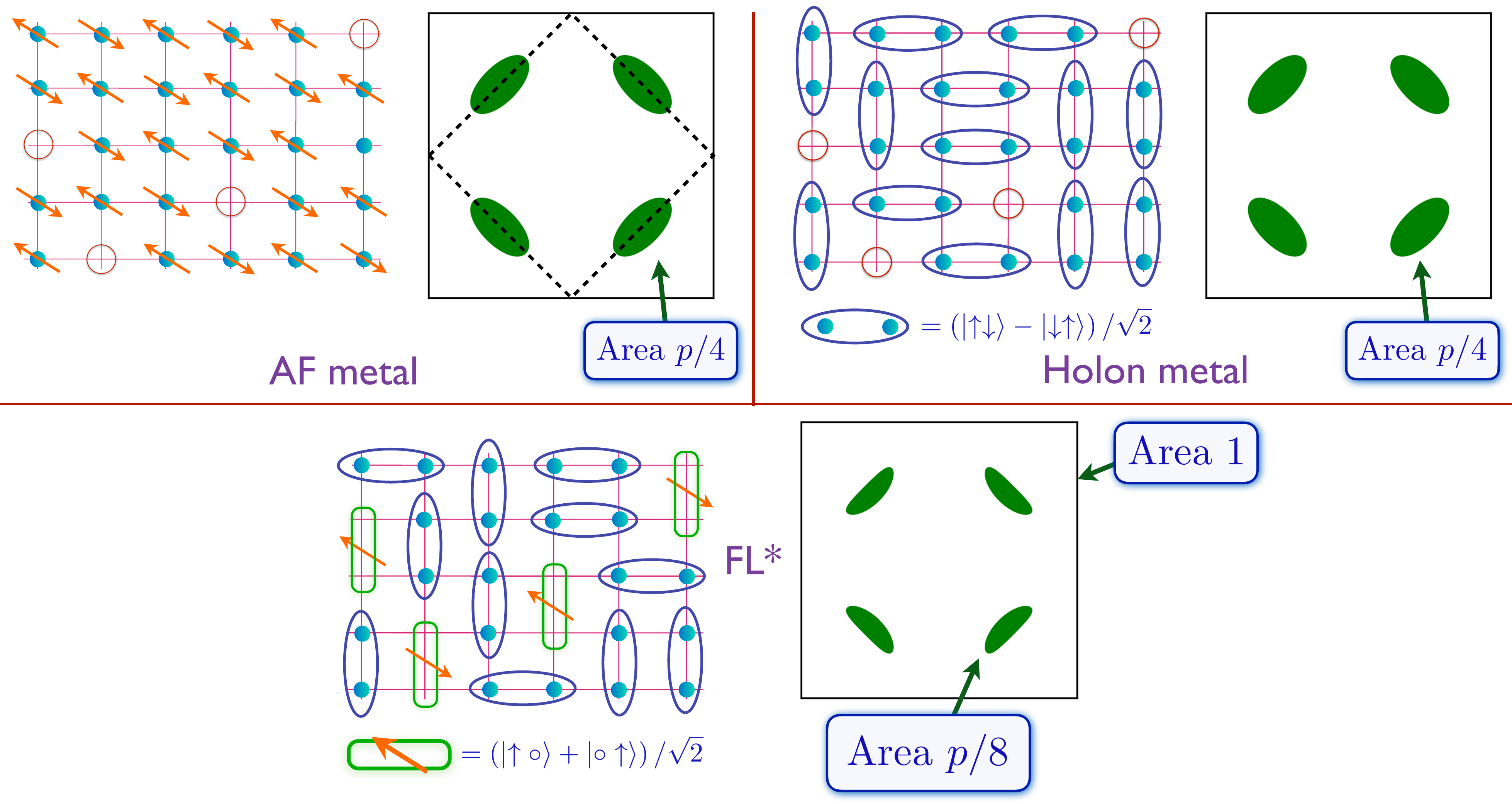}
\caption{Cartoon pictures of different states of doped antiferromagnets in a single band model with electron density $\rho=1-p$; adapted from Refs.~\cite{Punk15,ROPP25}. The areas, which are for hole-like Fermi surfaces, are those that would be measured by a probe such as quantum oscillations, and are specified in units in which the area of the Brillouin zone is unity. The AF metal has long-range antiferromagnetic order, and the reduced Brillouin zone is shown with the dashed line. The other phases do not break any symmetries. The open circles are holons, and these are assumed fermionic in the holon metal. The green dimers represent bound states of holons and spinons. Note that all 3 states would show a Hall coefficient corresponding to a density $p$ of positively charged carriers.}
\label{fig:metals}
\end{figure}

For the parton theories being analyzed here, we obtain the FL* state by also assuming $\mathcal{I}_{1z} = \mathcal{I}_{2} =0$ because of the stability of the parton perturbation theory. For $\mathcal{I}_{1p}$, we assume that the only pole contribution is from the emergent electron-like holon-spinon bound state $\widetilde{c}_\alpha$, and there is no contribution to $\mathcal{I}_{1p}$ from any of the spinons and holons. Then we immediately obtain
\beq
\rho-1 = 2 \times \mbox{(Area of $\widetilde{c}_\alpha$ Fermi surfaces)}
\eeq
{\it i.e.\/}
 the total area of the Fermi surfaces per spin is $(\rho-1)/2$ (in units where the area of the Brillouin zone is 1). We show a cartoon picture of the FL* state on a single-band square lattice model in Fig.~\ref{fig:metals}.
This picture employs a quantum dimer model formulation \cite{Punk15}, which can also be used to establish the area enclosed by the Fermi surfaces.

There can also be contributions to $\mathcal{I}_{1p}$ from the fermionic holons $h$, and we obtain the holon metal state \cite{Lee89,RKK08} if this is the only non-zero contribution to $\mathcal{I}_{1p}$ (and we have $\mathcal{I}_{1z} = \mathcal{I}_2 = 0$). If there are contributions to $\mathcal{I}_{1p}$ from both $\widetilde{c}_\alpha$ and $h$, then we obtain a hybrid state called the `holon-hole metal' \cite{RKK08}.

We note that bosonic quasiparticles, such as $B_{1,2}$ in Eqs.~(\ref{ll3},\ref{ll4}) cannot have zero frequency poles which contribute to $\mathcal{I}_{1p}$ \cite{Powell05}. Nevertheless Eq.~(\ref{ll4}) can be satisfied in a FL* state even though the l.h.s. has only boson number expectation values on the microscopic lattice scale---the contribution appears in the low energy theory from the appearance of spinon-holon bound states of $B_{1,2}$ and $f_\alpha$.

There are numerous non-perturbative analyses in the literature in which $\mathcal{I}_{1z}$ and/or $\mathcal{I}_2$ are non-zero \cite{Altshuler98,Dz03,Tsvelik06,YRZ,Kotliarzeros,Kane13,Scheurer18,Fabrizio22,Fabrizio22a,Si24,GleisKotliar24,Lucila1,Lucila2,P2_2025,GPS2,GKST,Shang21}.

The non-vanishing of $\mathcal{I}_2$ requires breakdown of integration by parts over frequency, and this is possible if the local density of states is divergent at zero frequency.
This happens for zero-dimensional models like the SYK model \cite{GPS2,GKST,Shang21}. 

A non-vanishing $\mathcal{I}_{1z}$ ({\it i.e.} a zero of the Green's function at $\omega=0$) is associated with a pole in the self-energy, which indicates the appearance of a new particle in some intermediate channel. This likely implies a breakdown of perturbation theory, which can be fixed by extending the set of renormalized fields $F_s$, so that the modified perturbative expansion has $\mathcal{I}_{1z} = 0$.
This improved theory (which should also have $\mathcal{I}_2 = 0$) will then provide a better handle on the true excitation spectrum of the state under consideration.

\FloatBarrier

\section{Torus flux insertions}
\label{sec:oshikawa}

Our discussion of the Luttinger constraints in Section~\ref{sec:gLutt} was perturbative, although the expansion was about parton saddle points which have non-trivial correlations built in. Here, we will sketch a non-perturbative approach to deriving these relations.
First, we will derive the Luttinger relation of the Fermi liquid as obtained by Oshikawa \cite{MO00}, and then turn in Sections~\ref{sec:Z2} and \ref{sec:dqcp} to the modified Luttinger relation (\ref{lutt2}) of the FL* phase. 

Consider an arbitrary quantum system, of bosons or fermions, 
defined on (say) a square lattice of unit lattice spacing, and placed on a torus.
The size of the lattice is $L_x \times L_y$, and we impose periodic boundary conditions. Assume the system has a global U(1)
symmetry, and all the local operators carry integer U(1) charges. 
Pick an eigenstate of the Hamiltonian (usually the ground state) $\left| \Psi (0) \right\rangle$.
Because of the translational symmetry, this state will obey
\beq
\hat{T}_x \left| \Psi (0) \right\rangle = e^{i P_x} \left| \Psi (0) \right\rangle,
\eeq
where $\hat{T}_x$ is the lattice translational operator by one lattice spacing along the $x$ direction, 
and $P_x$ is the momentum of the state $\left| \Psi (0) \right\rangle$.
Note that $P_x$ is only defined modulo $ 2\pi$. The state $\left| \Psi (0) \right\rangle$ will also have definite total U(1) charge, which we denote
by the integer $N$. 

\begin{figure}
\begin{center}
\includegraphics[height=4cm]{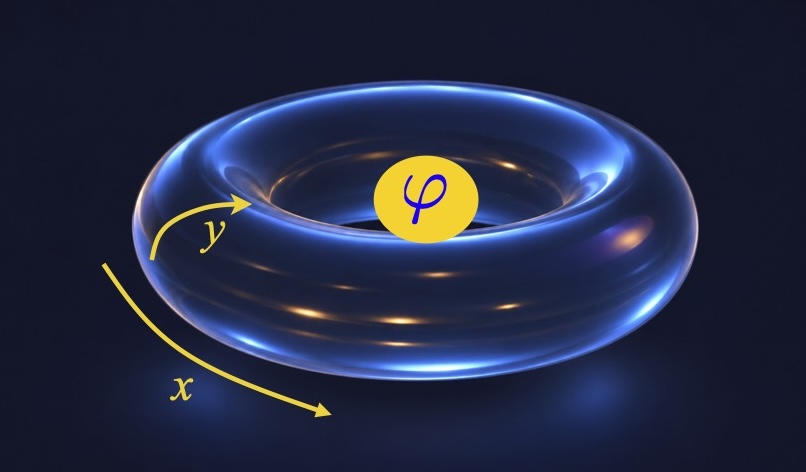}
\end{center}
\caption{Torus geometry with a flux quantum $\varphi$ of a global U(1) symmetry inserted.}
\label{fig:torusf}
\end{figure}

Now we gauge the global U(1) symmetry, and insert one flux quantum (with flux $2 \pi$) through one of the cycles of the torus over a long time $\mathcal{T}$
(see Fig.~\ref{fig:torusf}).
The initial and final Hamiltonians of the flux insertion process are related by a gauge transformation
\beq
\mathcal{U}_g H (\mathcal{T}) \mathcal{U}_g^{-1} = H (0) \quad , \quad \mathcal{U}_g = \exp \left( i \frac{2 \pi}{L_x} \sum_i x_i \hat{n}_i \right)\,,
\label{oshi2}
\eeq
where $\hat{n}_i$ is the integer number operator of the U(1) symmetry,
while the wavefunction evolves from $\left| \Psi (0) \right\rangle$ to $ \mathcal{U}_\mathcal{T} \left| \Psi (0) \right\rangle$,
where $\mathcal{U}_\mathcal{T}$ is the time evolution operator. We want to work in a fixed gauge in which the initial
and final Hamiltonians are the same: in this gauge, the final state is 
\begin{equation}
\left| \Psi (\mathcal{T} ) \right\rangle = \mathcal{U}_g  \, \mathcal{U}_\mathcal{T} \left| \Psi (0) \right\rangle. 
\end{equation}
Then the crystal momentum of $\left| \Psi (\mathcal{T} ) \right\rangle$ will differ from $P_x$ by $\Delta P_x$ with
\beq
\Delta P_x = \frac{2 \pi}{L_x} N \, (\mbox{mod} \, 2 \pi).
\label{deltaP}
\eeq
This follows from the easily established properties
\beq
\hat{T}_x \, \mathcal{U}_\mathcal{T} = \mathcal{U}_\mathcal{T} \,\hat{T}_x  \quad , \quad \hat{T}_x \,\mathcal{U}_g  = 
\exp \left( - i 2 \pi \frac{N}{L_x} \right) \mathcal{U}_g \,\hat{T}_x\,,
\eeq
up to an overall sign of $\Delta P_x$, which depends on the orientation conventions for $\mathcal{U}_g$ and $\hat T_x$, and plays no role in what follows.
Writing $N = \nu L_x L_y$, where $\nu$ is the U(1) charge per site, Eq.~(\ref{deltaP}) becomes
\begin{align}
\Delta P_x = 2 \pi \nu L_y \, (\mbox{mod $2\pi$}).
\label{deltaPa}
\end{align}

\subsection{Fermi liquid}
\label{sec:fl}

The discussion so far has been quite general, and not specified anything about the many-body system, apart from
its translational invariance and global U(1) symmetry. In the subsequent discussion, we will make further assumptions
about the nature of the ground state and low-lying excitations, and compute $\Delta P_x$ by other methods. Equating such a result
to (\ref{deltaP}) will then lead to important constraints on the allowed structure of the many-body ground state.

First, we assume the ground state is a Fermi liquid.
So its only low-lying excitations are fermionic quasiparticles around the Fermi surface. For our subsequent discussion,
it is important to also include the electron spin index, $\alpha = \uparrow, \downarrow$, 
and so we will have a Fermi liquid with 2 global U(1) symmetries, associated
respectively with the conservation of electron number and the $z$-component of the total spin, $S^z$. Consequently, there will  be two Luttinger
theorems, one for each global U(1) symmetry. The action for the fermionic quasiparticles, $c_{{\bm k}\alpha}$, with dispersion 
$\varepsilon ({\bm k})$ is
\beq
\mathcal{S}_{FL} = \int d\tau \int \frac{d^2 k}{4 \pi^2} \sum_{\alpha = \pm 1} 
c_{{\bm k} \alpha}^\dagger \left( \frac{\partial}{\partial \tau} 
- \frac{i}{2} \alpha A_\tau^s - i A_\tau^e  + \varepsilon ({\bm k} - \alpha {\bm A}^s /2 - {\bm A}^e) \right) c_{{\bm k} \alpha} \, ,
\label{SFL}
\eeq
where $\tau$ is imaginary time; in the gauge coupling $\alpha = \pm 1$. The Fermi surface is defined by $\varepsilon ({\bm k}) = 0$, and $\mathcal{S}_{FL}$ only applies for ${\bm k}$
near the Fermi surface, although we have (for notational convenience) written it in terms of an integral over all ${\bm k}$.
We have also coupled the quasiparticles to 2 probe gauge fields $A_\mu^e = (A_\tau^e, {\bm A}^e)$ and 
$A_\mu^s = (A_\tau^s, {\bm A}^s)$ which couple to the 2 conserved U(1) currents associated, respectively, with the conservation of electron
number and $S^z$.

We place the Fermi liquid on a torus, and insert a $2 \pi$ flux of a gauge field that couples only to the up-spin electrons.
So we choose $A_\mu^s = 2 A_\mu^e \equiv A_\mu$. Then the general momentum balance in (\ref{deltaP}) requires that
\beq
\Delta P_x = \frac{2 \pi}{L_x} N_\uparrow \, (\mbox{mod} \, 2 \pi) = \frac{2 \pi}{L_x} \frac{N}{2} \, (\mbox{mod} \, 2 \pi),
\label{deltaP2}
\eeq
where we assume equal numbers of up and down spin electrons $N_\uparrow = N_\downarrow = N/2$.

Now we compute the momentum balance {\em assuming} that the only low energy excitations are quasiparticles near the Fermi surface described by $\mathcal{S}_{FL}$, and these react like free particles to a sufficiently slow flux insertion. 
So each quasiparticle
picks up a momentum 
\beq 
\delta {\bm p} \equiv \left(\frac{2 \pi}{L_x} , 0 \right),
\eeq
and then we can write (with $\delta n_p $ the quasiparticle density excited by the flux insertion)
\begin{eqnarray}
\Delta P_x &=& \sum_p \delta n_p p_x .\nonumber
\end{eqnarray}
Now $\delta n_p = \pm 1$ on a shell of thickness $\delta {\bm p} \cdot d {\bm S}_p$ on the Fermi surface (where $d {\bm S}_p$ is an area element on the Fermi surface). So we can write the
above as a surface integral
\begin{eqnarray}
\Delta P_x
&=& \oint_{{\rm FS}} p_x \left( \frac{L_x L_y}{4 \pi^2} \right) \delta {\bm p} \cdot d {\bm S}_p \nonumber \\
&=& (\delta {\bm p} \cdot \hat{x}) \int_{{\rm FV}} \left( \frac{L_x L_y}{4 \pi^2} \right) d^2 p \nonumber
\end{eqnarray}
by the divergence theorem. So
\beq
\Delta P_x = \frac{2 \pi}{L_x} \left( L_x L_y \frac{V_{\rm FS}}{4 \pi^2} \right) \,,
\label{Vfs}
\eeq
where $V_{\rm FS}$ is the momentum space area enclosed by the Fermi surface; the factor within the brackets on the right-hand-side equals
the number of momentum space points inside the Fermi surface. Note that the entire contribution to the right-hand-side of (\ref{Vfs}) comes
from the vicinity of the Fermi surface where the quasiparticles are well-defined; we have merely used a mathematical identity to convert
the result to equal the volume, and we are not assuming the existence of quasiparticles far from the Fermi surface.

Now we equate (\ref{deltaP2}) and (\ref{Vfs}), along with a corresponding argument along the $y$ direction, and obtain
\beq
N_\uparrow - L_x L_y \frac{V_{\rm FS}}{4 \pi^2} = L_x m_x \quad, \quad N_\uparrow - L_x L_y \frac{V_{\rm FS}}{4 \pi^2} = L_y m_y
\eeq
for some integers $m_x$, $m_y$. By choosing $L_x$, $L_y$ mutually prime integers we can now show \cite{MO00,APAV04}
\beq
\frac{N_\uparrow}{L_x L_y} = \frac{V_{\rm FS}}{4 \pi^2} + m
\eeq
for some integer $m$.
This is the Luttinger relation for a Fermi liquid, stated in Eq.~(\ref{lutt1}) for $N_\uparrow = N_\downarrow = N/2$.

\subsection{FL* with a $\mathbb{Z}_2$ spin liquid}
\label{sec:Z2}

We now apply the same flux insertion to a FL* state. As in Section~\ref{sec:fl}, we insert a $2\pi$ flux of the gauge field $A_\mu$ that couples only to up-spin electrons, $A_\mu^s = 2 A_\mu^e \equiv A_\mu$. The exact result (\ref{deltaP2}) still holds, and with $N_\uparrow = N/2 = \rho L_x L_y/2$ it reads
\beq
\Delta P_x = 2 \pi \, \frac{\rho}{2} \, L_y \, (\mbox{mod} \, 2 \pi)\,. \label{deltaP3}
\eeq
The FL* state has two sectors of low-energy degrees of freedom, as we discussed in Section~\ref{sec:gLutt}:
\begin{enumerate}
\item Fermi surfaces of the quasiparticles $\widetilde{c}_\alpha$. These are gauge-neutral bound states of a spinon and a holon, and they carry the charge and spin of an electron. They respond to the flux exactly as the quasiparticles of Section~\ref{sec:fl}, and contribute the momentum (\ref{Vfs}).
\item A spin liquid, whose excitations are spinons carrying emergent gauge charge, together with the emergent gauge field. We write its contribution to the momentum change as (following (\ref{deltaPa}))
\beq
\Delta P_x^{\rm SL} = 2 \pi \nus L_y \, (\mbox{mod} \, 2 \pi)\,, \label{delta}
\eeq
which defines $\nus$.
\end{enumerate}
Equating the sum of the two contributions to (\ref{deltaP3}), and repeating the argument with mutually prime $L_x$, $L_y$ as in Section~\ref{sec:fl}, we obtain
\beq
\frac{\rho}{2} = \frac{V_{\rm FS}}{4 \pi^2} + \nus \, (\mbox{mod} \, 1)\,. \label{eq:master}
\eeq
For a Fermi liquid $\nus = 0$, and we recover the Luttinger relation (\ref{lutt1}). For the FL* relation (\ref{lutt2}), we need $\nus = 1/2$.

So the task is to show that the spin liquid absorbs momentum $\pi L_y$ under the flux insertion. It may seem surprising that a spin liquid can absorb momentum at all, since it has an energy gap. If its ground state on the torus were unique, the slow flux insertion would return it to the same state with the same momentum, and so $\nus$ would vanish. This is the content of the Lieb--Schultz--Mattis--Oshikawa--Hastings theorem \cite{LSM61,LSMOH1,LSMOH2}: an insulator with a half-odd-integer spin per unit cell cannot have a gap and a unique, symmetric ground state. A topologically ordered spin liquid evades this theorem because it has several degenerate ground states on the torus. The flux insertion can move the system from one of these to another with a different momentum. We now show that this is precisely what happens in an `odd' $\mathbb{Z}_2$ spin liquid with a background $\mathbb{Z}_2$ gauge charge on each site, following Refs.~\cite{TSSSMV03,TSSSMV04,APAV04,Bonderson16}.

We use the Schwinger boson parton theory of Section~\ref{sec:gLutt}, with spinons $b_\alpha$, holons $h$, and the constraint (\ref{ll1}). The resulting U(1) gauge theory can be ``Higgsed'' down to $\mathbb{Z}_2$ by condensing a field which pairs spinons on nearby sites, and which therefore carries U(1) gauge charge 2 \cite{RS91}. The resulting $\mathbb{Z}_2$ spin liquid has the following excitations:
\begin{center}
\begin{tabular}{lll}
\toprule
excitation & description & spin \\
\midrule
$e$ & bosonic spinon $b_\alpha$ (unit $\mathbb{Z}_2$ gauge charge) & $S=\tfrac12$ \\
$m$ & vison ($\pi$ flux of the $\mathbb{Z}_2$ gauge field) & $S=0$ \\
$f=e\times m$ & fermionic spinon (bound state of $e$ and $m$) & $S=\tfrac12$ \\
\bottomrule
\end{tabular}
\end{center}
The key braiding property is that a particle with unit $\mathbb{Z}_2$ gauge charge ($e$ or $f$) acquires the Aharonov--Bohm phase $-1$ when it is taken around a vison. The holon $h$ also carries unit gauge charge, and so also acquires $-1$ around a vison, while the electron-like $\widetilde{c}_\alpha$ is gauge neutral.

\subsubsection{Step 1: the flux insertion creates a vison line}

We present the argument of Bonderson {\it et al.\/} \cite{Bonderson16}.

Consider first how the inserted flux acts on the gauge-charged partons. With $A^s_\mu = 2 A^e_\mu = A_\mu$, a particle with charge $q_e$ and spin $S^z$ couples to $A_\mu$ with charge $q_e/2 + S^z$; this equals 1 for an up-spin electron and 0 for a down-spin electron, as required. A spinon has $q_e=0$ and $S^z = \pm 1/2$, and a holon has $q_e = -1$ (see Eq.~(\ref{ll2})) and $S^z=0$. So every particle carrying unit gauge charge has charge $\pm 1/2$ under $A_\mu$, and acquires the phase $e^{\pm i \pi} = -1$ when it is taken around the $x$ cycle of the torus after the $2\pi$ flux has been inserted. The gauge-neutral $\widetilde{c}_\alpha$ quasiparticles have integer charge under $A_\mu$, and are simply boosted as in Section~\ref{sec:fl}.

A phase of $-1$ for every gauge-charged particle on the $x$ cycle is exactly the effect of a vison ($\pi$ flux) threaded through the hole of the torus.
\begin{figure}
\begin{center}
\includegraphics[height=4cm]{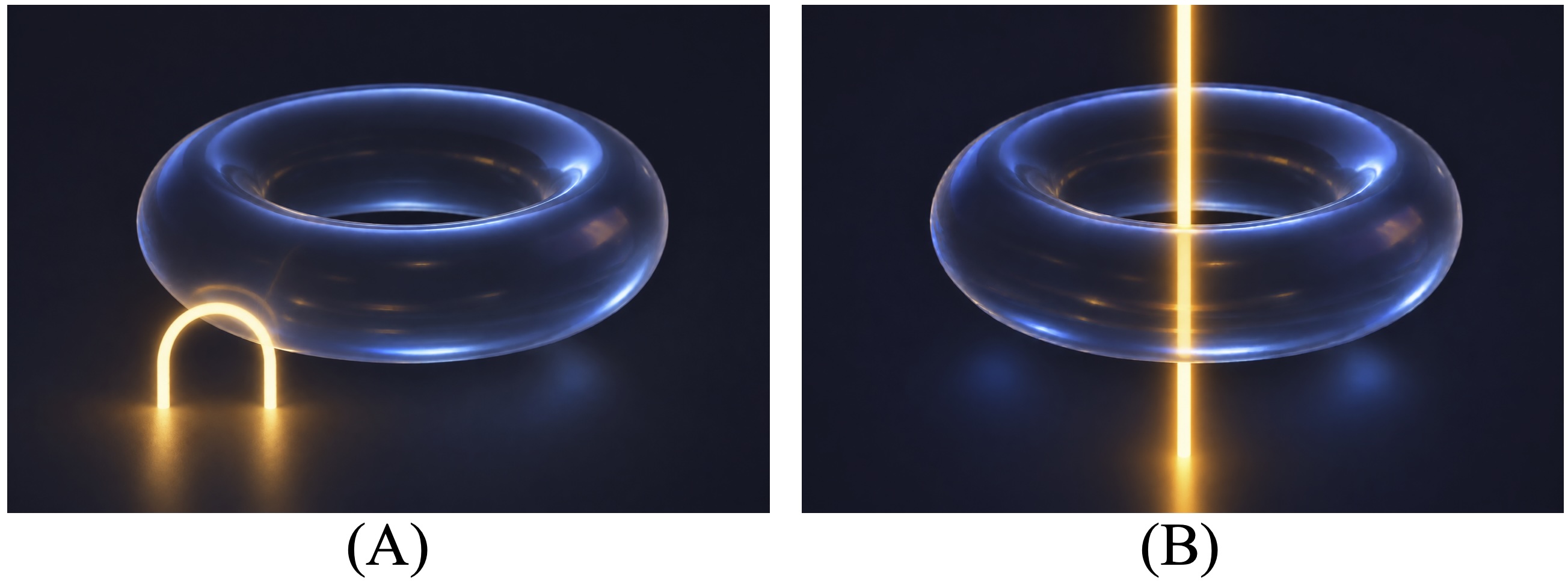}
\end{center}
\caption{(A) A pair of visons created by a $\pi$-flux tube. (B) We straighten out the flux tube into the center hole of the torus. This is topologically equivalent to moving the visons around the $y$ torus cycle (see Fig.~\ref{fig:torusf}) and annihilating them on the other side.}
\label{fig:torustube}
\end{figure}

\begin{figure}[h]
\centering
\begin{tikzpicture}[scale=1]
\draw[thick] (0,0) rectangle (6,4);
\foreach \x in {0.5,1.5,...,5.5} \foreach \y in {0.5,1.5,...,3.5}
  \fill[orange!80!black] (\x,\y) circle (3pt);
\draw[very thick,blue!70!black,dashed] (2,0) -- (2,4);
\draw[very thick,blue!70!black] (3,0) -- (3,4);
\draw[->,thick,blue!70!black] (2.1,4.35) -- (2.9,4.35);
\node[blue!70!black] at (2.5,4.85) {\small $\hat T_x$};
\node[left] at (0,2) {\small $y$};
\node[below] at (3,0) {\small $x$};
\node[right,align=left] at (7,2.6) {\small \textcolor{orange!80!black}{$\bullet$} = unit background gauge charge,\\[-1mm] \small \phantom{$\bullet$ =} Eq.~(\ref{ll1})};
\node[right] at (7,1.2) {\small \textcolor{blue!70!black}{line} = vison line $W_m$};
\end{tikzpicture}
\caption{After the flux insertion, the $\mathbb{Z}_2$ spin liquid contains a vison line running along $y$. Translating the line by one lattice spacing ($\hat T_x$; dashed: before, solid: after) drags it across a column of $L_y$ sites. Each site carries one unit of background gauge charge, which contributes the phase $-1$. The spin liquid therefore carries momentum $\pi L_y$.}
\label{fig:torus}
\end{figure}
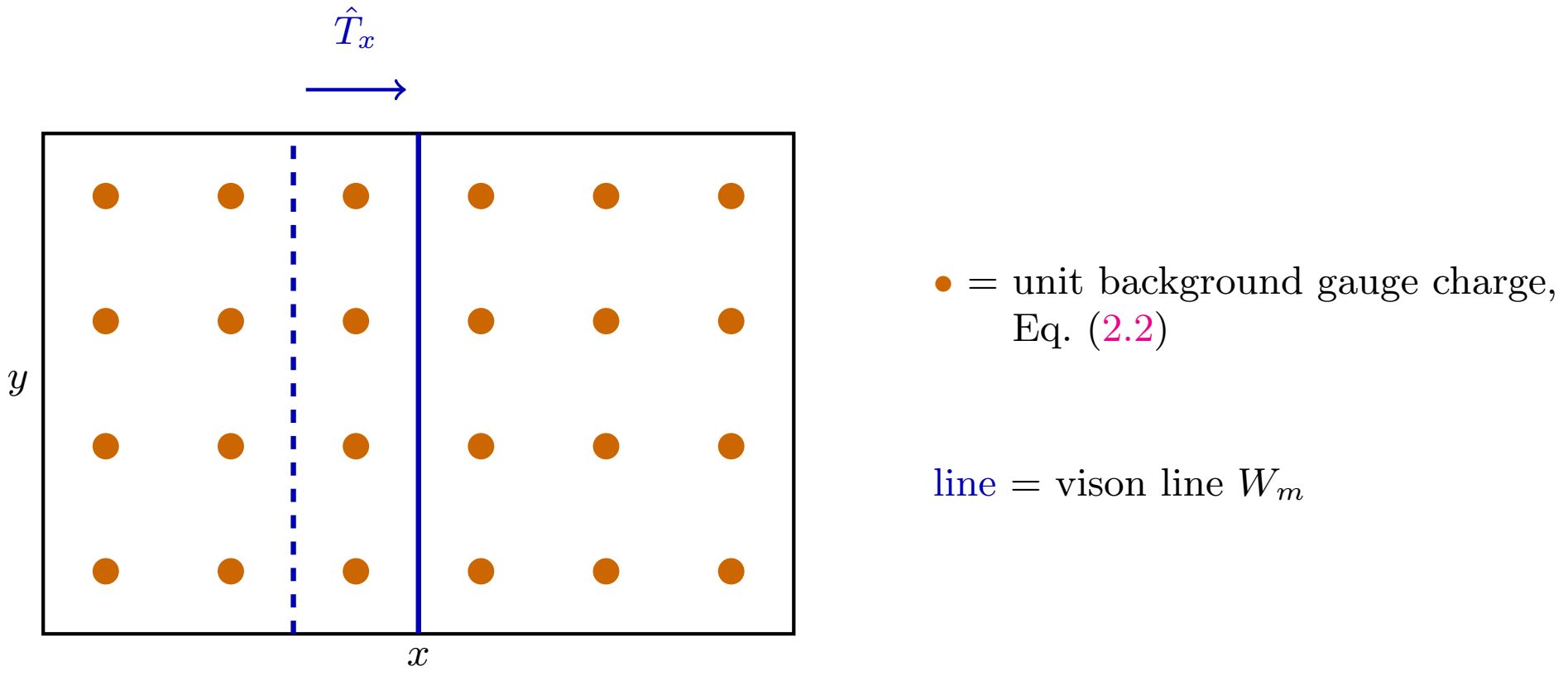

As shown in Fig.~\ref{fig:torustube}, threading a vison through the hole is topologically equivalent to creating a pair of visons, dragging one of them around the $y$ cycle, and annihilating the pair. This leaves behind a vison line running along the $y$ direction, which every $x$ loop must cross (Fig.~\ref{fig:torus}). We denote the operator that creates this line by $W_m$ (in the notation of Section 16.3 of Ref.~\cite{QPMbook}, this vison line operator is $V_y$). So, in the notation of Eq.~(\ref{oshi2}), the effect of the flux insertion on the spin liquid is
\beq
\mathcal{U}_g\, \mathcal{U}_\mathcal{T} \left| \Psi (0) \right\rangle = W_m \left| \Psi (0) \right\rangle\,, \label{vison}
\eeq
up to the boost of the $\widetilde{c}_\alpha$ Fermi surfaces, which we have already accounted for. Because the spin liquid is gapped, the flux insertion is adiabatic. The state $W_m \left| \Psi (0) \right\rangle$ is again a ground state on the torus, but in a different topological sector (the $\mathbb{Z}_2$ flux through the hole of the torus has changed) \cite{Bonderson16}.

\subsubsection{Step 2: the vison line carries momentum $\pi L_y$}

We now need the momentum of the state $W_m \left| \Psi (0) \right\rangle$ relative to $\left| \Psi (0) \right\rangle$. The key input is the constraint (\ref{ll1}): every site carries one unit of background gauge charge, independent of the doping. (As noted below Eq.~(\ref{ll1}), this is the origin of the Berry phase of $\pi$ for a vison encircling a site.) In the language of Ref.~\cite{Bonderson16}, there is a background anyon $e$ in every unit cell, and translations act projectively on a single vison \cite{RJSS91,SSMV99, HPS11,QPMbook},
\beq
\boxed{
\hat T_x \hat T_y \hat T_x^{-1} \hat T_y^{-1} = -1 \quad \mbox{on a vison.} } \label{visonPSG}
\eeq
This projective action of translations on visons, and its consequences, were analyzed in detail in Ref.~\cite{HPS11}: one consequence is that the vison states form degenerate multiplets related by translations, so a confinement transition driven by vison condensation must also break translation symmetry.

Translating the vison line by one lattice spacing in the $x$ direction sweeps it across one column of $L_y$ sites (Fig.~\ref{fig:torus}), and each site contributes a phase $-1$:
\beq
\hat T_x^{-1} W_m^{-1} \hat T_x W_m = (-1)^{L_y} = e^{2 \pi i \cdot \frac12 \cdot L_y}\,.
\eeq
So $W_m \left| \Psi (0) \right\rangle$ has momentum $\pi L_y$ relative to $\left| \Psi (0) \right\rangle$. Comparing with (\ref{delta}), we conclude
\beq
\nus = \frac{1}{2}\,,
\eeq
and Eq.~(\ref{eq:master}) gives
\beq
\frac{V_{\rm FS}}{4\pi^2} = \frac{\rho - 1}{2} \, (\mbox{mod} \, 1)\,,
\eeq
which is the FL* relation (\ref{lutt2}).

It is useful to check this result in the insulating spin liquid at $\rho=1$. Now there are no Fermi surfaces, and (\ref{deltaP3}) requires the spin liquid alone to absorb momentum $\pi L_y$. So the background anyon in each unit cell must acquire $-1$ when encircled by a vison. In the toric code, the only anyons with this property are $e$ and $f$. A $\mathbb{Z}_2$ spin liquid with no background anyon, or with a background vison $m$, would have $\nus = 0$; such spin liquids require an even number of spin-$1/2$ moments per unit cell \cite{Bonderson16}.

The value $\nus = 1/2$ was derived for a $\mathbb{Z}_2$ spin liquid decoupled from the Fermi surfaces. It continues to hold when the two sectors interact \cite{PCAS16}, for two reasons: the first follows from the vison energy gap, and the second from gauge invariance.
\begin{itemize}
\item $\nus$ is fixed by discrete topological data: the spinon carries spin $1/2$, and there is one unit of background gauge charge per site. These data cannot change continuously, and so cannot be changed by a weak coupling between the sectors, or by doping, without closing the vison gap.
\item The flux insertion acts on the two sectors in sharply different ways: gauge-charged particles see a vison line, while gauge-neutral quasiparticles are boosted. A $\widetilde{c}_\alpha$ quasiparticle cannot convert into a single spinon, because it is gauge neutral while the spinon carries gauge charge, a fact employed in Ref.~\cite{PCAS16}. Absorbing the background gauge charge into the Fermi sea requires a phase transition (confinement or Higgs condensation of the gauge field), and the result is the FL phase with $\nus = 0$.
\end{itemize}

\subsubsection{Interpretation as an anomaly}
\label{sec:Z2anomaly}

The result (\ref{deltaP}) is exact. It holds for every state and every Hamiltonian with the given symmetries, and it is fixed by the charge per unit cell. Any low-energy description of the system must therefore reproduce it. When a low-energy theory reproduces such a symmetry-protected response, which could not arise in a trivial, symmetric, gapped system, we say that the theory carries an \emph{'t~Hooft anomaly} for the symmetry. When the response comes from a lattice filling, as here, it is called a Lieb--Schultz--Mattis (LSM) anomaly, associated with the combination of the U(1) symmetry and translations \cite{CZBVB,Seiberg23,Meng26}.

In the language of Refs.~\cite{Metlitski18,CZBVB,CZ,SenthilElse21}, the LSM anomaly at filling $\nu$ (the U(1) charge per site) is encoded in a topological term
\beq
S_{\rm LSM}[\nu] = 2\pi i\,\nu\int_{X_4} x\cup y\cup \frac{dA}{2\pi}\,. \label{eq:SLSM}
\eeq
The meaning of the symbols is as follows.
\begin{itemize}
\item $A$ is a background (non-dynamical) gauge field for the U(1) symmetry, and $dA/2\pi$ is its flux density, normalized so that its integral over any closed surface is an integer number of flux quanta. For the argument of Section~\ref{sec:fl} we take $A$ to couple only to the up-spin electrons, so that $\nu = \rho/2$.
\item $x$ and $y$ are background ``gauge fields'' for the lattice translations $\hat T_x$ and $\hat T_y$ \cite{CZ}. They are integer-valued and live on the links of a (space-time) lattice; their integral along a closed curve counts the net number of translations encountered along it. On the $L_x \times L_y$ torus, the integral of $x$ around the $x$ cycle is $L_x$, and the integral of $y$ around the $y$ cycle is $L_y$. So $x$ and $y$ keep track of the system sizes, and a change in the size of the system is equivalent to inserting a ``flux'' of translations.
\item The cup product $\cup$ is the lattice analog of the wedge product of differential forms, so $x \cup y \cup dA/2\pi$ plays the role of a 4-form.
\item $X_4$ is a fictitious four-dimensional ``bulk'' whose boundary is the physical $2+1$ dimensional space-time. The physical theory is not invariant by itself under gauge transformations of the background fields $(x,y,A)$. The bulk term (\ref{eq:SLSM}) is chosen so that its change under these gauge transformations cancels that of the physical theory (``anomaly inflow'').
\end{itemize}
The reader can ignore the formal machinery, and read (\ref{eq:SLSM}) as a compact restatement of (\ref{deltaP}). In the flux insertion of Section~\ref{sec:oshikawa}, $\int dA/2\pi = 1$ on the surface spanned by time and the $x$ cycle (the threaded flux), $\int y = L_y$, and $x$ ``measures'' the change in the $\hat T_x$ eigenvalue. The phase $e^{2\pi i \nu L_y}$ produced by (\ref{eq:SLSM}) is precisely the momentum change $\Delta P_x = 2\pi\nu L_y$ of Eq.~(\ref{deltaP}).

The essential property of anomalies that we have used is that \emph{the anomalies of decoupled low-energy sectors add}, as in Eq.~(\ref{eq:master}). The Fermi surface contributes $V_{\rm FS}/(4 \pi^2)$, which is Oshikawa's computation in Section~\ref{sec:fl}, and the spin liquid contributes $\nus$. 

Equation (\ref{eq:master}) is useful only because $\nus$ is fixed. Otherwise, momentum could be transferred continuously between the sectors and we could conclude nothing about $V_{\rm FS}$. We needed two facts about the spin liquid:
\begin{enumerate}
\item[(Q)] \textbf{Quantization:} $\nus$ takes values in a discrete set, and cannot change under weak perturbations, including the coupling to the Fermi surfaces.
\item[(V)] \textbf{Value:} for the spin liquid of interest, $\nus=1/2$.
\end{enumerate}
This is the precise sense in which the FL* shift of $\rho$ in Eq.~(\ref{lutt2}) is quantized. The volume $V_{\rm FS}$ varies continuously with doping, but its deficit relative to the Luttinger value (\ref{lutt1}) is pinned to the quantized anomaly of the spin liquid.

\subsection{FL* with a deconfined critical spin liquid}
\label{sec:dqcp}

For the $\mathbb{Z}_2$ spin liquid, the quantization (Q) followed from the rigidity of topological order, and the value (V) from the background gauge charge in each unit cell \cite{RJSS91,SSMV99,HPS11}. We now ask whether a FL* can also be built on a gapless spin liquid with a continuous gauge symmetry. Now there is no gap to make the flux insertion adiabatic, there are no topologically degenerate ground states, and no gap protects $\nus$ against the coupling to the Fermi surfaces. We will see that the essential steps of Section~\ref{sec:Z2} nevertheless go through, with the role of the gap played by spin-rotation symmetry and the quantized LSM anomaly of the gapless theory \cite{Metlitski18}. This gapless case plays a central role in recent applications of FL* to the cuprates \cite{ROPP25,Seamus26}.

We consider the best-studied gapless spin liquid in two dimensions: the critical state at the deconfined quantum critical point (DQCP) between N\'eel and valence-bond-solid (VBS) order in the $S=1/2$ square lattice antiferromagnet \cite{SVBSF,SBSVF}. The resulting FL* is a metal in which the $\widetilde{c}_\alpha$ Fermi surfaces coexist with this critical spin liquid. Because the DQCP is a critical point, this FL* is a critical metal rather than a stable phase. Numerical studies indicate that the N\'eel--VBS transition is either continuous or very weakly first order \cite{Yasir19,Nahum:2015vka,Becca20,Imada21,Fuzzy24,Meng24,Gu24,Chester24,Sandvik24}; the anomaly arguments below do not depend on which is the case.

\subsubsection{The $\mathbb{CP}^1$ theory and its symmetries}

We keep the Schwinger boson spinons of Section~\ref{sec:Z2}, but do not Higgs the U(1) gauge field. Near the N\'eel--VBS transition, the low-energy theory of the spin liquid is the $\mathbb{CP}^1$ model \cite{SVBSF,SBSVF}
\beq
\mathcal{L}_{\mathbb{CP}^1} = |(\partial_\mu - i a_\mu) z_\alpha|^2 + s\,|z_\alpha|^2 + u\,(|z_\alpha|^2)^2 + \ldots\,, \label{cp1}
\eeq
where $z_\alpha$ ($\alpha = \uparrow,\downarrow$) is the continuum limit of the spinon $b_\alpha$: a boson with spin $S=1/2$ and unit charge under the emergent U(1) gauge field $a_\mu$. The N\'eel order parameter is $\vec{n} \sim z^\dagger \vec{\sigma} z$. The VBS order parameter is the monopole operator $V$, which inserts $2\pi$ flux of $a_\mu$ \cite{NRSS89prl,NRSS90}. For $s < s_c$ the spinons condense, giving N\'eel order; for $s > s_c$ the monopoles proliferate, giving VBS order. At the critical point $s = s_c$, the spinons, the photon, and the monopoles are all gapless.

The continuum theory (\ref{cp1}) has three global symmetries:
\begin{itemize}
\item SO(3)$_s$ spin rotations, under which $z_\alpha$ transforms as a spinor. Gauge-invariant operators, such as $\vec{n}$, carry integer spin, so SO(3), rather than SU(2), acts faithfully on them.
\item A ``topological'' or ``magnetic'' U(1)$_\Phi$ symmetry, whose conserved current is $\epsilon^{\mu\nu\lambda} \partial_\nu a_\lambda/(2\pi)$: the conserved charge is the total flux of $a$. The monopole $V$ has unit charge, $V \rightarrow e^{i\alpha} V$.
\item A unitary ``charge conjugation'' $C$: $z \rightarrow i \sigma^y z^\ast$, $a \rightarrow -a$, $V \rightarrow V^\dagger$.
\end{itemize}

Next, we turn to the representation of the $S=1/2$ square lattice antiferromagnet by the continuum theory $\mathcal{L}_{\mathbb{CP}^1}$. U(1)$_\Phi$ and $C$ together form the group O(2)$_\Phi$ of rotations and reflections of the two-component VBS order parameter $({\rm Re}\,V, {\rm Im}\,V)$. The microscopic lattice symmetries act in the continuum theory as elements of O(2)$_\Phi$, combined with continuum translations and rotations \cite{Zheng89,senthil2,Metlitski18}:
\beq
\hat T_x:~ V \rightarrow V^\dagger\,, \quad \hat T_y:~ V \rightarrow - V^\dagger\,, \quad \hat R_{\pi/2}:~ V \rightarrow i V\,,  \label{tcp1}
\eeq
with $\hat T_x$ also acting as $C$ on $z$ and $a$. So we have the key relations
\begin{align}
\boxed{
\hat T_x = C \quad , \quad  \hat T_y = U^\Phi_\pi C, 
} \label{CPTxTy}
\end{align}
where $U^\Phi_\pi$ is a rotation by $\pi$ in U(1)$_\Phi$. These relations encode the analog of Eq.~(\ref{visonPSG}) for the $\mathbb{Z}_2$ spin liquid; see Eq.~(\ref{TTdirect}) below.
These unusual transformations are the continuum remnant of the Berry phases of the spins: the background gauge charge on each site in Eq.~(\ref{ll1}) gives monopoles a Berry phase which depends on the plaquette on which they are located \cite{Haldane88,NRSS89prl,NRSS90}. They are also the reason that $V$ is the order parameter of the square lattice VBS. Note that the diagonal translation $\hat T_x \hat T_y$ acts as $V \rightarrow -V$: a monopole carries crystal momentum.

The lattice symmetries (\ref{tcp1}) do not forbid the quadrupled monopole term $V^4 + V^{\dagger 4}$, which breaks the continuous U(1)$_\Phi$ down to $\mathbb{Z}_4$. In the VBS phase ($s > s_c$) this term is relevant: it pins the VBS order to a columnar or plaquette pattern, which breaks translation symmetry and enlarges the unit cell, and so changes the Luttinger constraint. At the critical point it is believed to be (dangerously) irrelevant. Importantly, the arguments below do not require the full U(1)$_\Phi$ symmetry. They use only $C$ and the rotation $U^\Phi_\pi$ in Eq.~(\ref{CPTxTy}), which belong to the group generated by $C$ and $\mathbb{Z}_4$ that the lattice preserves. So the $V^4$ term does not affect them; we need only assume that the system is at the critical point, and not in the VBS phase.

\subsubsection{The dual Dirac fermion description} 

Wang {\it et al.\/} \cite{Wang17} proposed that the same critical point is described by a dual theory of $N_f=2$ flavors of two-component Dirac fermions coupled to an emergent SU(2) gauge field (``QCD$_3$''). This is the low-energy theory of the $\pi$-flux state of fermionic spinons on the square lattice. In this description the global symmetry is enlarged to SO(5), which rotates the five-component vector $(\vec{n}, {\rm Re}\,V, {\rm Im}\,V)$ made of the N\'eel and VBS order parameters; the fermions transform as spinors of SO(5). Such an emergent SO(5) symmetry has been observed numerically at the DQCP \cite{Nahum:2015vka}. The Dirac fermion theory carries the same anomaly as the $\mathbb{CP}^1$ theory (discussed below), in the form $\pi i \int w_4[{\rm SO}(5)]$, where $w_4$ is the fourth Stiefel--Whitney class of the background SO(5) gauge field \cite{Wang17}. Restricting SO(5) to the symmetries visible in the $\mathbb{CP}^1$ theory reproduces the continuum anomaly (\ref{cp1anom0}) in Section~\ref{sec:cp1anom} below \cite{Metlitski18}. The lattice anomaly (\ref{cp1anom1}) requires additional microscopic input, which is not contained in either field theory: the way the lattice symmetries are embedded in the continuum symmetry group, Eq.~(\ref{tcp1}). This embedding is the same in both descriptions \cite{Wang17}. On the SO(5) vector $(\vec{n}, {\rm Re}\,V, {\rm Im}\,V)$, $\hat T_x$ acts as ${\rm diag}(-1,-1,-1,1,-1)$ and $\hat T_y$ as ${\rm diag}(-1,-1,-1,-1,1)$: the N\'eel vector is reversed, and the VBS order parameter is reflected as in Eq.~(\ref{tcp1}). Because the two theories have the same continuum anomaly and the same embedding of the lattice symmetries, all the conclusions below hold in either description. We use the $\mathbb{CP}^1$ language, which connects directly to the Schwinger boson description of Section~\ref{sec:Z2}.

\subsubsection{Flux insertion}
\label{sec:direct}

We now show that the continuum $\mathbb{CP}^1$ theory, supplemented only by the action (\ref{tcp1}) of the lattice translations, yields $\Delta P_x^{\rm SL}$ and $\nus$. The continuum theory contains no lattice and no background gauge charge, so Step 2 of Section~\ref{sec:Z2}, which swept a vison line across the background charges, cannot be repeated directly. Instead, we replace the flux through the hole of the torus by an exactly equivalent geometry, in which the $2\pi$ flux of $A_\mu$ pierces the torus itself. The momentum balance then becomes a statement about the commutation of $\hat T_x$ and $\hat T_y$, which Eq.~(\ref{CPTxTy}) determines.

\emph{(a) Uniform flux and magnetic translations.} Consider any lattice system with a conserved U(1) charge $N = \nu L_x L_y$, and place a uniform total flux $2\pi$ of $A_\mu$ through the $L_x \times L_y$ torus. We use the Landau gauge $A_x = -2\pi y/(L_x L_y)$, $A_y = 0$. The translation $\hat T_x$ still commutes with the Hamiltonian. A translation by one lattice spacing in $y$ shifts $A_x$, and so it must be combined with a gauge transformation of $A_\mu$ (a ``magnetic translation''). In this subsection, $\hat T_y$ denotes this combined operation. A charge transported around a unit cell encloses the flux $2\pi/(L_x L_y)$, and so
\beq
\hat T_x\, \hat T_y\, \hat T_x^{-1}\, \hat T_y^{-1} = e^{-2\pi i N/(L_x L_y)} = e^{-2\pi i \nu}\,. \label{magTT}
\eeq
This uniform-flux geometry carries the same information as the flux through the hole. Translating by $L_y$ steps in $y$ shifts $A_x$ by $2\pi/L_x$, which is one flux quantum through the $x$ cycle. So $\hat T_y^{L_y}$ is the large gauge transformation $\mathcal{U}_g$ of Eq.~(\ref{oshi2}), and the $L_y$-th power of Eq.~(\ref{magTT}) is Oshikawa's relation $\hat T_x\, \mathcal{U}_g = e^{-2\pi i N/L_x}\, \mathcal{U}_g\, \hat T_x$, i.e.\ Eq.~(\ref{deltaPa}). The same holds sector by sector, since the sectors are decoupled and their charges add, as in Eq.~(\ref{eq:master}). So the momentum absorbed by the spin liquid in Eq.~(\ref{delta}) can be read off from its projective phase in uniform flux:
\beq
\hat T_x\, \hat T_y\, \hat T_x^{-1}\, \hat T_y^{-1}\Big|_{\rm SL} = e^{-2\pi i \nus} \quad \Longleftrightarrow \quad \Delta P_x^{\rm SL} = 2\pi \nus L_y\,. \label{magTTSL}
\eeq

\emph{(b) The emergent flux is half-integer.} The spin liquid is electrically neutral, and couples to $A_\mu = A^s_\mu$ only through $S^z$. So $z_\uparrow$ couples to $a_\mu + A_\mu/2$, and $z_\downarrow$ to $a_\mu - A_\mu/2$. Each of these must be a properly quantized U(1) gauge field, with integer total flux through the torus. Since $\int dA/2\pi = 1$, this requires
\beq
\int \frac{da}{2\pi} \in \mathbb{Z} + \frac{1}{2}\,, \quad \mbox{i.e.} \quad Q_\Phi \in \mathbb{Z} + \frac{1}{2}\,, \label{halfQ}
\eeq
where $Q_\Phi$ is the U(1)$_\Phi$ charge of Section~\ref{sec:dqcp} (the total flux of $a_\mu$ in units of $2\pi$, so that $V$ has $Q_\Phi = 1$). In other words, a $2\pi$ flux of spin rotations acts on the spinons as a $\pi$ flux of the emergent gauge field, just as, in Section~\ref{sec:Z2}, it acted as a vison. Every state of the spin liquid in this background has half-odd-integer U(1)$_\Phi$ charge.

\emph{(c) Eq.~(\ref{CPTxTy}) gives the projective phase.} In the continuum theory, the translations act as in Eq.~(\ref{CPTxTy}), now combined with gauge transformations of $A_\mu$. The continuum fields carry no net $S^z$, so these gauge transformations contribute no phase, and the commutator is determined by O(2)$_\Phi$ alone. A reflection reverses the sense of a rotation, $C\, U^\Phi_\theta\, C = U^\Phi_{-\theta}$, so Eq.~(\ref{CPTxTy}) gives
\beq
\hat T_x\, \hat T_y\, \hat T_x^{-1}\, \hat T_y^{-1}\Big|_{\rm SL} = C\, U^\Phi_\pi\, C\, U^\Phi_{-\pi} = U^\Phi_{-2\pi} = e^{-2\pi i Q_\Phi} = -1\,, \label{TTdirect}
\eeq
where the last step uses Eq.~(\ref{halfQ}). Comparing with Eq.~(\ref{magTTSL}),
\beq
\nus = \frac{1}{2}\,, \qquad \Delta P_x^{\rm SL} = \pi L_y \pmod{2\pi}\,. \label{nusdirect}
\eeq

The two inputs have a transparent physical meaning. Eq.~(\ref{halfQ}) states that the spinons carry spin $1/2$. Eq.~(\ref{tcp1}) encodes the background gauge charge on each lattice site through the monopole Berry phases. Without the latter, the translations would commute with U(1)$_\Phi$, and the phase in Eq.~(\ref{TTdirect}) would be 1. The phase $-1$ in Eq.~(\ref{TTdirect}) is the U(1) counterpart of the projective action (\ref{visonPSG}) of translations on a vison: the $\pi$ flux of the emergent gauge field induced by the spin flux sees a phase $-1$ per unit cell, like the vison. The same computation gives the expected answers in two other cases:
\begin{itemize}
\item For $S=1$ on the square lattice, the monopole Berry phases make both translations act as the same reflection, $\hat T_x = \hat T_y = C$ \cite{Metlitski18}. The commutator in Eq.~(\ref{TTdirect}) is then $C^4 = 1$, so $\nus = 0$.
\item On the honeycomb lattice with $S=1/2$ per site, the N\'eel--VBS transition (now to a Kekul\'e VBS) is described by the same $\mathbb{CP}^1$ theory, but the translations $\hat T_{1,2}$ along the two lattice vectors act as rotations, $\hat T_1 = U^\Phi_{2\pi/3}$ and $\hat T_2 = U^\Phi_{-2\pi/3}$ \cite{Metlitski18}. These commute, so $\nus = 0$ (mod 1), as appropriate for two spin-$1/2$ moments per unit cell. In this case the flux insertion argument gives no protected deficit in the Fermi volume.
\end{itemize}

\subsubsection{The LSM anomaly of the $\mathbb{CP}^1$ theory}
\label{sec:cp1anom}

Section~\ref{sec:direct} showed that $\nus = 1/2$ follows from the action (\ref{tcp1}) of the lattice translations, even though the continuum theory (\ref{cp1}) contains no lattice and no explicit background charge. We now place this result in the general framework of the LSM anomaly of the $\mathbb{CP}^1$ theory, the analog of Eq.~(\ref{eq:SLSM}) for the $\mathbb{Z}_2$ spin liquid (Section~\ref{sec:Z2anomaly}). We first describe the anomaly of the continuum theory, and then its restriction to the lattice symmetries. Section~\ref{sec:nuscp1} uses the result to derive $\nus$.

First, we ignore the lattice, and consider the full internal symmetry O(2)$_\Phi \times$ SO(3)$_s$ of the $\mathbb{CP}^1$ theory (\ref{cp1}). Its anomaly was computed by Komargodski {\it et al.\/} \cite{KSTZ19}, and rederived in Ref.~\cite{Metlitski18}:
\beq
S_{\rm cont} = \pi i \int_{X_4} w_2[\xi_\Phi] \cup \Big( w_2^s + w_1[\xi_\Phi] \cup w_1[\xi_\Phi] \Big)\,. \label{cp1anom0}
\eeq
Here $\xi_\Phi$ is a background gauge field for O(2)$_\Phi$, and $w_1[\xi_\Phi]$, $w_2[\xi_\Phi]$ are $\mathbb{Z}_2$-valued fields which characterize it, in the same way as $w_2^s$ characterizes the SO(3)$_s$ background:
\begin{itemize}
\item $w_2^s$ is the second Stiefel--Whitney class of the SO(3)$_s$ background gauge field. It is a $\mathbb{Z}_2$-valued (0 or 1) ``flux'' of the spin rotations: it equals 1 (mod 2) on a surface pierced by $2\pi$ flux of spin rotations, which a spin-$1/2$ particle sees as a phase $-1$, and an integer-spin particle does not see at all. It is the SO(3)$_s$ analog of $dA/2\pi$ in Eq.~(\ref{eq:SLSM}).
\item $w_1[\xi_\Phi]$ lives on the links and records the reflections. Its integral around a closed curve is 1 (mod 2) if, along that curve, the background acts on $({\rm Re}\,V, {\rm Im}\,V)$ by a reflection (an O(2) element of determinant $-1$, such as $C$).
\item $w_2[\xi_\Phi]$ is the $\mathbb{Z}_2$ ``flux'' of the O(2)$_\Phi$ background. For a background of U(1)$_\Phi$ rotations only, with gauge field $A_\Phi$, it is $dA_\Phi/2\pi$ (mod 2). A $2\pi$ flux of $A_\Phi$ is the background-field version of a vortex in the VBS order parameter, so $w_2[\xi_\Phi]$ counts VBS vortices mod 2.
\end{itemize}
The two terms obtained by expanding the bracket in Eq.~(\ref{cp1anom0}) have a simple meaning. The first term is the field-theory form of the Levin--Senthil result \cite{LevinSenthil04}. In the presence of a VBS vortex ($w_2[\xi_\Phi] = 1$), what remains is $\pi i \int w_2^s$, the response of a Haldane chain, whose end carries spin $1/2$. So a VBS vortex binds a spin $1/2$. The second term arises because $C$ reverses the N\'eel vector, $\vec{n} \rightarrow -\vec{n}$, as follows from $z \rightarrow i \sigma^y z^\ast$. So, wherever the O(2)$_\Phi$ background contains a reflection, the spin background is effectively an O(3) rather than an SO(3) background, and its $\mathbb{Z}_2$ flux becomes $w_2^s + w_1[\xi_\Phi] \cup w_1[\xi_\Phi]$ \cite{Metlitski18}.

Now we consider the consequences of the lattice for the continuum $\mathbb{CP}^1$ theory. Metlitski and Thorngren \cite{Metlitski18} determined the anomaly of the $\mathbb{CP}^1$ theory directly in terms of the microscopic symmetries of the square lattice: the SO(3)$_s$ spin rotations and the lattice translations $\hat T_x$, $\hat T_y$. The idea is the same as in the flux-threading argument. Making $L_x$ or $L_y$ odd is equivalent to twisting the boundary conditions by a translation, and by Eq.~(\ref{tcp1}) such a twist acts in the continuum theory as a reflection of the VBS order parameter, $V \rightarrow \pm V^\dagger$. One then asks how the continuum theory responds to these twists, combined with twists by spin rotations. Their result, whose derivation from Eqs.~(\ref{cp1anom0}) and (\ref{tcp1}) we describe below, is
\beq
S_{\mathbb{CP}^1} = \pi i \int_{X_4} \Big( x \cup y \cup w_2^s + x^3 \cup y + x \cup y^3 \Big)\,. \label{cp1anom1}
\eeq
As in Eq.~(\ref{eq:SLSM}), $X_4$ is a fictitious four-dimensional bulk whose boundary is the physical space-time, $\cup$ is the lattice analog of the wedge product, and $x^3 \equiv x \cup x \cup x$. The other symbols are:
\begin{itemize}
\item $x$ and $y$ are the translation gauge fields of Eq.~(\ref{eq:SLSM}). Their integrals count translations, so that $\oint x = L_x$ around the $x$ cycle of the torus and $\oint y = L_y$ around the $y$ cycle. Because of the factor $\pi i$, only the parity matters: $x$ and $y$ enter Eq.~(\ref{cp1anom1}) only mod 2, and an odd $L_x$ (or $L_y$) acts as a single twist by $\hat T_x$ (or $\hat T_y$).
\end{itemize}
Following Ref.~\cite{Metlitski18}, we distinguish two kinds of anomalies. An \emph{intrinsic} anomaly is a property of the microscopic symmetries themselves, as they act on the lattice. It must be reproduced by every low-energy description of the lattice model, and it forbids a trivial symmetric gapped state however strong the perturbations; the LSM anomaly is the prototype. An \emph{emergent} anomaly is an anomaly of the symmetries as they act in the low-energy theory, which may differ from their microscopic action. Here, for example, the translations form the group $\mathbb{Z}$ on the lattice, but by Eq.~(\ref{tcp1}) they act on $V$ as $\mathbb{Z}_2$ reflections. An emergent anomaly constrains only the vicinity of the critical point, and can be removed by sufficiently strong perturbations. The two parts of Eq.~(\ref{cp1anom1}) are of these two kinds:
\begin{itemize}
\item The first term of Eq.~(\ref{cp1anom1}), $\pi i \int x \cup y \cup w_2^s$, is the \emph{intrinsic} LSM anomaly. It remains non-trivial when $x$ and $y$ are integer-valued, as appropriate for the microscopic translations, and it is a genuine, discrete anomaly. It is precisely the response of a square lattice with spin $1/2$ per unit cell. To see this, view the lattice as the surface of a three-dimensional stack of Haldane spin chains, running perpendicular to the surface; each chain ends in a spin $1/2$ on the surface. Making $L_y$ odd ($\oint y = 1$) reduces the bulk action to $\pi i \int x \cup w_2^s$, which is the response of a one-dimensional array, along $x$, of Haldane chains: increasing $L_y$ by one toggles each column between a trivial chain and a Haldane chain \cite{Metlitski18}. Section~\ref{sec:nuscp1} shows how this term determines $\nus$.

The same term can be read in two complementary ways. If we insert a $2\pi$ flux of spin rotations ($w_2^s = 1$), what remains is $\pi i \int x \cup y$: the lattice translations act projectively, with $\hat T_x \hat T_y \hat T_x^{-1} \hat T_y^{-1} = -1$, and the spin liquid absorbs momentum $\pi$ per row, i.e.\ $\nus = 1/2$. This is the direct argument of Section~\ref{sec:direct}. The second reading is the Haldane-chain response obtained above with $L_y$ odd. It is the lattice counterpart of the Levin--Senthil result contained in the first term of Eq.~(\ref{cp1anom0}): Metlitski and Thorngren showed that the analogous term for the $90^\circ$ rotations of the square lattice implies that a rotation-symmetric VBS vortex carries spin $1/2$ \cite{Metlitski18}. So the spin $1/2$ in the core of a VBS vortex and the Luttinger volume deficit of the DQCP FL* are consequences of a single anomaly, which records the spin $1/2$ per unit cell of the lattice.
\item The last two terms of Eq.~(\ref{cp1anom1}) are \emph{emergent} anomalies. They vanish when $x$ and $y$ are integer-valued, because then $x \cup x = 0$ (mod 2). They exist only because, by Eq.~(\ref{tcp1}), the translations act as $\mathbb{Z}_2$ reflections of the VBS order parameter in the continuum theory ($\hat T_x^2$ and $\hat T_y^2$ act trivially on $V$), which is a property of the vicinity of the critical point rather than of the lattice. They are the response of a three-dimensional $\mathbb{Z}_2 \times \mathbb{Z}_2$ symmetry-protected phase, and imply that no weak translation-invariant perturbation of the critical spin liquid can open a trivial gap, even one which breaks spin-rotation symmetry; a sufficiently strong staggering of the exchange couplings, however, does lead to a trivial gapped state \cite{Metlitski18}. They do not involve the spin rotations or the U(1) symmetry, and so do not constrain the Fermi volume.
\end{itemize}

It is worth contrasting the two anomalies (\ref{eq:SLSM}) and (\ref{cp1anom1}). Strictly speaking, a fractional U(1) filling is an ``anomaly-like'' relation between the microscopic and low-energy theories rather than a genuine 't~Hooft anomaly: the coefficient $\nu$ in Eq.~(\ref{eq:SLSM}) can be varied continuously, and so that term can be deformed continuously to zero \cite{SenthilElse21}. By contrast, the coefficient of $\pi i \int x \cup y \cup w_2^s$ can only be 0 or 1, and this is a genuine, discrete anomaly. Its discreteness is what quantizes $\nus$ in the gapless spin liquid.

Finally, we present the promised derivation of the lattice anomaly (\ref{cp1anom1}) from the continuum anomaly (\ref{cp1anom0}) and the action (\ref{tcp1}) of the lattice translations \cite{Metlitski18}. In terms of the two real components of $V$, Eq.~(\ref{tcp1}) reads
\beq
\hat T_x:~ ({\rm Re}\,V, {\rm Im}\,V) \rightarrow ({\rm Re}\,V, -{\rm Im}\,V)\,, \qquad
\hat T_y:~ ({\rm Re}\,V, {\rm Im}\,V) \rightarrow (-{\rm Re}\,V, {\rm Im}\,V)\,. \label{TxTyReIm}
\eeq
So ${\rm Im}\,V$ is odd only under $\hat T_x$, and ${\rm Re}\,V$ is odd only under $\hat T_y$. In a background of translation twists, ${\rm Im}\,V$ therefore sees a sign twist given by $x$ (mod 2), and ${\rm Re}\,V$ sees a sign twist given by $y$ (mod 2). An O(2) background of this form is a sum of two independent sign twists. Its Stiefel--Whitney classes then follow from the product rule $1 + w_1 + w_2 = (1 + x)(1 + y)$ (the Whitney sum formula):
\beq
w_1[\xi_\Phi] = x + y\,, \qquad w_2[\xi_\Phi] = x \cup y \pmod 2\,. \label{Whitney}
\eeq
The second relation states that a $\hat T_x$ twist around one cycle of a torus, combined with a $\hat T_y$ twist around the other, has the same effect on the $\mathbb{Z}_2$ flux as a single VBS vortex. We insert Eq.~(\ref{Whitney}) into Eq.~(\ref{cp1anom0}), and use $(x+y) \cup (x+y) = x \cup x + y \cup y$ (mod 2); the cross terms cancel because cup products of these $\mathbb{Z}_2$ fields commute mod 2. This gives
\beq
S_{\rm cont} \;\rightarrow\; \pi i \int_{X_4} x \cup y \cup \big( w_2^s + x \cup x + y \cup y \big)\,,
\eeq
which is Eq.~(\ref{cp1anom1}).

The intrinsic term $x \cup y \cup w_2^s$ is therefore the product of two facts: a VBS vortex binds a spin $1/2$ (the first term of Eq.~(\ref{cp1anom0})), and, by Eq.~(\ref{tcp1}), a pair of $\hat T_x$ and $\hat T_y$ twists acts as a VBS vortex (the second relation in Eq.~(\ref{Whitney})). The second fact relies on $\hat T_x$ and $\hat T_y$ acting as two \emph{different} reflections, as in Eq.~(\ref{CPTxTy}). For the two other cases considered in Section~\ref{sec:direct}, $w_2[\xi_\Phi] = 0$ and the intrinsic anomaly vanishes, in agreement with $\nus = 0$ found there. For the honeycomb lattice the entire anomaly (\ref{cp1anom0}), including its emergent part, vanishes \cite{Metlitski18}; this also permits the tripled monopole $V^3 + V^{\dagger 3}$ as a perturbation of the critical theory.

\subsubsection{From the LSM anomaly to the momentum balance}
\label{sec:nuscp1}

We now show explicitly how the intrinsic term of Eq.~(\ref{cp1anom1}) determines the momentum $\Delta P_x^{\rm SL}$ absorbed by the spin liquid, defined in Eq.~(\ref{delta}). The emergent terms do not involve the spin rotations and play no role here.

\emph{(i) Restrict to $S^z$ rotations.} The flux insertion of Section~\ref{sec:fl} uses only spin rotations about the $z$ axis, with background gauge field $A^s$ coupling to $S^z$. The SO(3)$_s$ background is then a U(1) background, and its $\mathbb{Z}_2$ flux $w_2^s$ becomes the U(1) flux mod 2:
\beq
w_2^s \;\rightarrow\; \frac{dA^s}{2\pi} \pmod 2\,. \label{w2reduce}
\eeq
This expresses the fact that a $2\pi$ rotation about the $z$ axis gives the phase $e^{2\pi i S^z} = -1$ to a spin-$1/2$. The spin liquid is electrically neutral, so it couples to $A_\mu = A^s_\mu = 2A^e_\mu$ only through $A^s_\mu$. The intrinsic term therefore becomes
\beq
\pi i \int_{X_4} x \cup y \cup w_2^s \;\rightarrow\; \pi i \int_{X_4} x \cup y \cup \frac{dA}{2\pi} = 2\pi i \cdot \frac{1}{2} \int_{X_4} x \cup y \cup \frac{dA}{2\pi} = S_{\rm LSM}[1/2]\,, \label{cp1toLSM}
\eeq
which is the LSM anomaly (\ref{eq:SLSM}) at filling $\nu = 1/2$.

\emph{(ii) Reduce on the $y$ cycle.} On the $L_x \times L_y$ torus, $\oint y = L_y$. Integrating over the $y$ cycle leaves a term in one lower dimension:
\beq
S_{\rm LSM}[1/2] \;\rightarrow\; 2\pi i \, \frac{L_y}{2} \int_{X_3} x \cup \frac{dA}{2\pi}\,. \label{cp1reduce}
\eeq
This is the LSM anomaly of a one-dimensional system along $x$, whose unit cell is a column of $L_y$ sites, carrying total charge $q = L_y/2$ under $A_\mu$.

\emph{(iii) Thread the flux.} For a one-dimensional ring with charge $q$ per unit cell, the term $2\pi i\, q \int x \cup dA/2\pi$ is the one-dimensional analog of Eq.~(\ref{deltaP}): threading $2\pi$ flux of $A$ through the ring ($\int dA/2\pi = 1$ on the surface spanned by the $x$ cycle and time) multiplies the $\hat T_x$ eigenvalue by $e^{2\pi i q}$. With $q = L_y/2$ this gives
\beq
e^{i \Delta P_x^{\rm SL}} = e^{2\pi i\, L_y/2} \quad \Longrightarrow \quad \Delta P_x^{\rm SL} = \pi L_y = 2\pi \cdot \frac{1}{2} \cdot L_y \pmod{2\pi}\,. \label{cp1DeltaP}
\eeq
Comparing with Eq.~(\ref{delta}), we obtain $\nus = 1/2$. For $L_y$ odd, the $\hat T_x$ eigenvalue of the critical spin liquid flips sign. This agrees with the direct argument of Section~\ref{sec:direct}, Eq.~(\ref{nusdirect}).

\subsubsection{Quantization of $\nus$ without a gap}

We can now check the conditions (Q) and (V) of Section~\ref{sec:Z2} for the deconfined critical spin liquid.
\begin{itemize}
\item[(V)] $\nus = 1/2$ follows from the direct argument, Eq.~(\ref{nusdirect}), or equivalently from the intrinsic anomaly in Eq.~(\ref{cp1anom1}). Neither argument needs a gap. Eqs.~(\ref{halfQ}) and (\ref{TTdirect}) are operator statements. They hold in every state of the $\mathbb{CP}^1$ theory, gapped or gapless, and rely only on the spin $1/2$ of the spinons and on Eq.~(\ref{CPTxTy}). The difference from the $\mathbb{Z}_2$ case is that the flux of the emergent gauge field is now a fluctuating variable rather than a topological label. So the flux insertion generally takes the ground state to an excited state, at an energy of order $1/L$, rather than to another ground state. Oshikawa's momentum bookkeeping needs only the momentum.
\item[(Q)] In the direct argument, the projective phase (\ref{TTdirect}) can only be $\pm 1$, because $Q_\Phi$ is either an integer or a half-odd-integer; so $\nus \in \{0, 1/2\}$. In the language of the LSM anomaly, the spin liquid is electrically neutral, so its only response to the flux of $A$ comes from $S^z$ rotations, which are part of SO(3)$_s$. With SO(3)$_s$ symmetry, the coefficient of $x \cup y \cup w_2^s$ can only be 0 or 1: the moments in a unit cell carry either integer or half-odd-integer spin, and there is no third option. A discrete quantity cannot change continuously, and so a weak coupling to the Fermi surfaces, which preserves the total SO(3)$_s$, cannot change $\nus$ unless the low-energy theory itself is reorganized (for example, at a transition to the FL phase). The quantization of the monopole Berry phases upon doping was also discussed in Refs.~\cite{BS07,PCAS16}.
\end{itemize}
This is the gapless counterpart of the topological protection in the $\mathbb{Z}_2$ case. It is also its limitation: in a Zeeman field, SO(3)$_s$ is broken to U(1), the critical spin liquid acquires a continuously variable magnetization, and its share $\nus$ of the up-spin filling is no longer pinned. This is the distinction noted in Section~\ref{sec:cp1anom}: the U(1) filling anomaly can be deformed continuously, while the SO(3)$_s$ anomaly $\pi i\int x\cup y\cup w_2^s$ cannot.

\subsection*{Acknowledgements}
This research was supported by the US National Science Foundation under Grant DMR-2245246. Section~\ref{sec:gLutt} has appeared previously in an unpublished article on the arXiv \cite{SSlectures25}. Section~\ref{sec:fl} has been adapted from Ref.~\cite{QPMbook}. Sections~\ref{sec:Z2} and \ref{sec:dqcp} were written with the assistance of numerous interactive sessions with Anthropic's Claude Opus 5.5.

\bibliography{refs}

\end{document}